\documentclass[a4paper,10pt]{article}

\usepackage{graphicx}

\title{Superconductivity in a two dimensional extended Hubbard model}
\author{
E. J. Calegari and S. G. Magalh\~aes\footnote{ggarcia@ccne.ufsm.br} \\ \\
{\it Laborat\'orio de Mec\^anica Estat\'{\i}stica e Teoria da Mat\'eria Condensada}\\
{\it Universidade Federal de Santa Maria, 97105-900 Santa Maria, RS, Brazil}\\  \\
A. A. Gomes \\ \\
{\it Centro Brasileiro de Pesquisas F\'{\i}sicas-CBPF}\\
{\it  Rua Xavier Sigaud 150, 22290-180, Rio de Janeiro, RJ, Brazil}}

\begin{document}

\maketitle

\begin{abstract}
The Roth's two-pole approximation has been used by the present authors
to investigate the role of $d-p$ hybridization in the superconducting
properties of an extended $d-p$ Hubbard model. Superconductivity with
singlet $d_{x^2-y^2}$-wave pairing is treated by following Beenen and Edwards formalism.
In this work, the Coulomb interaction, the temperature and the
superconductivity have been considered in the calculation of some relevant
correlation functions present in the Roth's band shift.
The behavior of the order parameter associated with temperature, hybridization,
Coulomb interaction and the Roth's band shift effects on superconductivity are
studied.
\end{abstract}

\twocolumn
\section{Introduction}
\label{intro}
After  almost two decade of intense research about the cuprates, there is still plenty of open
questions in this problem. However, it is recognized
that the electrons which move on the $CuO_2$
planes are the most relevant to describe their physical properties \cite{ref4}.
In the undoped regime, these compounds are insulators and
exhibit antiferromagnetic order at sufficient low temperatures
\cite{ref4,ref3}. The physical properties of the insulating phase
can be well described by the Heisenberg model \cite{ref3}. Upon
doping, these systems suppress the
antiferromagnetic order and become superconductors. In this
scenario there is no doubt that the $d$-$d$ electron correlations play a fundamental role.

The study of the electronic structure near the Fermi level $\varepsilon_F$
in such strongly correlated systems is very important to understand their physical properties
\cite{ref1}. Earlier angle-resolved photoemission experiments (ARPES) have
showed the presence of flat bands close to $\varepsilon_F$ in a region centered around the
point $(\pi,0)$ in the $p$-type cuprates
like $Bi_2$$Sr_2$$CuO_6$ and $YBa_2$$Cu_3$$O_y$ \cite{ref1,ref2}.
Due to the presence of strong correlations, to study some physical properties of these
cuprate compounds, the one-band Hubbard model \cite{ref5} can be used.
Bulut {\it et al.} \cite{ref6,ref7} have done Monte Carlo calculations
in the one-band Hubbard model. Their results show bands with a flat
region near $(\pi,0)$ point for a given doping
which agreed with the previously mentioned ARPES results \cite{ref1}.

Beenen and Edwards \cite{ref9}, using the Roth's two-pole
approximation \cite{ref8} in the one-band Hubbard model, have studied
the normal state of the model obtaining flat quasi-particle bands,
which agree well with those found with Monte Carlo
simulations \cite{ref6,ref7}. 
The Roth's two-pole approximation has been proposed to improve the Hubbard-I approximation
\cite{ref5} by considering a decoupling scheme which 
produces an additional
energy shift (the Roth's band shift) in the peaks of the spectral function. 
That result is in agreement
with those obtained by Harris and Lange \cite{ref8.1}. 
They looked at the moments 
of individual peaks in the spectral function. The presence of the exchange term $\langle S_iS_j\rangle$
in the Roth's band shift exhibits in it a spin dependence. As consequence, the Roth's 
method raises the possibility of magnetic solution in the Hubbard model while this feature is
not present in the Hubbard-I approximation.
Recently, due to the good agreement between the Roth's and the
Monte Carlo data, Beenen and Edwards have extended the Roth's two-pole
approximation in order to investigate the superconducting properties of the
one-band Hubbard model. Their main achievement has been to show
the emergence of the pairing with $d_{x^{2}-y^{2}}$ symmetry in a
given amount of doping.
In that approach, the  gap equation for $d$-wave symmetry depends
on a particular four operator correlation function which, in
principle, can be found extending the Roth's formalism to obtain
two particle Green's functions. However,  the authors have
introduced two decoupling schemes to calculate the gap. The first
one (the factorization procedure) has been formulated to treat the
problem for intermediated values of $U$ (the Cou\-lomb interaction)
and it provides an upper estimate for the gap and $T_{c}$ (the
critical temperature). The second one is adapted for very large
$U$ scenario which preserves the proper limit for
$U\rightarrow\infty$ where the gap function vanishes.

Nevertheless, the one-band models neglect the presence of the
oxygen sites. Due to the strong correlations at the $Cu$-sites,
the oxygen sites may be occupied by holes when the system is doped
\cite{ref3}. For instance, the Hubbard one-band model suffers some
limitations to describe the low-energy physical properties of the
cuprate superconductors \cite{ref5.1}. In the doped regime, the
one-band Hubbard model gives a wrong description of various
properties, like, for example, the asymmetric magnetic
doping-temperature phase diagram \cite{ref5.2}. Therefore, a model
which take into account also the oxygen can be more adequate to
treat the cuprate systems in the doped regime \cite{Emery}. This
raises the question whether it is possible to extend the Beenen and
Edwards analysis to investigate the $d$-wave symmetry
superconductivity when the hybridization is present.

Recently, the present authors have used the extended Hubbard model \cite{Emery}
with the Roth's method to study the role of hybridization in the superconductivity
following closely the approach introduced by Beenen and Edwards \cite{ref11}.
As discussed in the references \cite{ref9} and \cite{ref10}, the
flattening of the bands is directly related to the band shift.
The presence of flat bands at Fermi level $\varepsilon_F$ in the $p$-type cuprates
\cite{ref1} suggests a high density of states at the Fermi level,
which can favor pair formation. Therefore,
considering that the main responsible elements for the density of states at the Fermi level are the $d$-electrons,
it has been assumed in Ref.\cite{ref11} that the $d-d$ pairs are the most relevant ones for superconductivity \cite{ref13}.

The band shift plays an important role in the study of the
superconducting properties of the model using the Roth's or some
similar procedures. In reference \cite{ref11}, the factorization
procedure \cite{ref9} has been used to investigate the effects of
the hybridization on the superconductivity. It has been shown
\cite{ref11} that the hybridization has strong effects in the
shift and, therefore, in some superconducting physical properties
such as the critical temperature $T_{c}$. However, as a first
approach, in reference \cite{ref11}, the band shift has been
evaluated taking into account the hybridization effects, but
disregarding temperature effects, superconducting properties and,
most important, it  has been  considered in the limit
$U\rightarrow \infty$. As a consequence of this limit, many
 correlation functions, which appear in the shift, are vanished.
The important point is that these correlation functions are
very relevant in the sense of to include correctly the hybridization effects. 
Therefore, it
would  be necessary to calculate the shift with the $U$ finite in
order to include the hybridization effects in a more complete way.

In this work, the superconductivity problem has been  studied
using the Roth's method, following closely reference
\cite{ref9}, but adapted to the $d$-$p$ extended Hubbard model.
Here, special attention is devoted to the effects of
the hybridization and superconductivity in the band shift.
In order to have the effects of the hybridization  included properly in the
superconductivity, the gap function is obtained using the factorization procedure \cite{ref11} and
the shift is evaluated with finite $U$.
This procedure is justified because it preserves some correlation functions
present in the band shift, which are non vanishing for finite $U$.
As consequence, it captures the effects of the hybridization properly.
Some preliminary results of this approach have been given in Ref. \cite{ref12}.

There are some shortcomings in the Beenen and Edwards approach
\cite{ref14,Avella}. For instance,  the $d_{x^{2}-y^{2}}$ pairing
is quite dependent on the choice of the decoupling scheme for the
correlation functions related to the gap. However, in the present work,
the main goal is to study the effects of hybridization.
Therefore, as discussed in the previous paragraph, the natural choice
is the decoupling scheme for intermediated $U$, which is also the simplest one.
One is allowed to find in that procedure, at least, a better
estimate for the gap (and therefore for $T_{c}$)
as a function of hybridization
within the same decoupling procedure.

The paper presents the following organization. In section
\ref{sec:2} it is introduced the model and given a short
introduction of the Roth's method \cite{ref8}. Also, some analytic
expressions for quasi-particle bands and the Green's functions are
derived. In section \ref{sec:3}, the factorization procedure
proposed by Beenen and Edwards \cite{ref9} is applied for the
present case. In section \ref{sec:4}, the band shift is discussed
in detail. The numerical results are showed and discussed in
section \ref{sec:5}. Finally, in section \ref{sec:6}, a
short summary and some concluding remarks are given.

\section{General formulation}
\label{sec:2}

The model considered here assumes overlapping bands. It is
characterized by a narrow $d$-like band with a large density of
states and a wide $p$-like band with low density of states. The
extended Hubbard model is defined as:
\begin{eqnarray}
H&=&\sum_{i,\sigma }(\varepsilon _{d}-\mu)d_{i\sigma }^{\dag}d_{i\sigma
}+\sum_{i,j,\sigma }t_{ij}^{d}d_{i\sigma }^{\dag}d_{j\sigma }+
U\sum_{i}n_{i\uparrow}^{d}n_{i\downarrow}^{d}\nonumber\\
 & &+\sum_{i,\sigma }(\varepsilon _{p}-\mu)p_{i\sigma }^{\dag}p_{i\sigma
}+\sum_{i,j,\sigma }t_{ij}^{p}p_{i\sigma }^{\dag}p_{j\sigma }\nonumber\\
 & &+\sum_{i,j,\sigma }t_{ij}^{pd}\left( d_{i\sigma
}^{\dag}p_{j\sigma +}p_{i\sigma }^{\dag}d_{j\sigma }\right)
\label{eq2.0}
\end{eqnarray}
where $\mu$ is the chemical potential. The $d_{i\sigma }^{\dag}$, $d_{i\sigma}$
and $p_{i\sigma }^{\dag}$, $p_{i\sigma}$
are the creation and annihilation operators of the $d$- and $p$-electrons,
respectively, with spin $\sigma$ at a lattice site $i$. The $\varepsilon_{d}$ and
$\varepsilon_{p}$ are the centers of the on site energies of the occupied orbitals of the
copper and oxygen respectively. The second term of the
Hamiltonian given in Eq. (\ref{eq2.0}) describes a narrow
$d$-band with a hopping amplitude $t^d$. The Hamiltonian
(\ref{eq2.0}) considers also a $p$-band which is wider than the
$d$-band. The following relation between $t^d$ and $t^p$ can be
established $t^p=\alpha t^d$ with $\alpha > 1$. Also, $t^d < 0$ to
coincide the bottom of the $d$- and $p$-bands with the $\Gamma$ point
$k_x=k_y=0$ as suggested by experimental results \cite{ref1}. The third term corresponds to the
Coulomb interaction $U$ that represents the repulsion between two holes in the same $d$-orbital.
The last term of the Hamiltonian (\ref{eq2.0}) is the $d-p$
hybridization and describes the nearest neighbor hopping
process between the $d$-orbital of the Cu-atom and the $p$-orbital
of the O-atom. Considering a rectangular two dimensional lattice, the unperturbed
$d$- and $p$-energy bands are given by
\begin{equation}
\varepsilon_{\vec{k}}^d = 2t^d(\cos(k_xa)+\cos(k_ya))
\label{eq2.01}
\end{equation}
and
\begin{equation}
\varepsilon_{\vec{k}}^p = 2t^p(\cos(k_xa)+\cos(k_ya))
\label{eq2.02}
\end{equation}
where $a$ is the lattice constant.

In this work, the Hamiltonian given in Eq.
(\ref{eq2.0}) has been investigated using the Roth's two-pole
approximation \cite{ref8} to obtain the Green's function in the
Zubarev's formalism. In the Roth's procedure, a set of operators $\left\{A_{n}\right\}$
is introduced in order to describe the relevant one particle excitations of the system.  
These operators satisfy in some approximation the following relation:
\begin{equation}
\left[ A_{n},H\right] _{\left( -\right) }=\sum_{m}K_{nm}A_{m}
\label{eq2.1}.
\end{equation}

Anticommuting both sides of Eq. (\ref{eq2.1}) with each operator of the set
$\left\{A_{n}\right\}$ and taking the thermal average, the equation (\ref{eq2.1})
becomes:
\begin{equation}
E_{nm}=\sum_mK_{nm}N_{nm}
\label{eq2.2}
\end{equation}
where $E_{nm}$ and $N_{nm}$ are the energy and normalization matrices, given by
\begin{equation}
E_{nm}=\left\langle\left[ \left[ A_{n},H\right] _{\left( -\right) },
A_{m}^{\dag}\right] _{\left( +\right) }\right\rangle
\label{eq2.3}
\end{equation}
and
\begin{equation}
N_{nm}=\langle [ A_{n},A_{m}^{\dag}]_{\left( +\right) }\rangle
\label{eq2.4}.
\end{equation}
In matrix notation, Eq. (\ref{eq2.2}) is written as
$\bf{E}=\bf{K\cdot N}$, where, if $\bf N$ is nonsingular, then the $\bf
K$ matrix can be obtained. With the equation of motion (in the
Zubarev's formalism) of the Green's function
\begin{equation}
G_{nm}\left( \omega \right) =\langle\langle A_n;A_{m}^{\dagger}\rangle\rangle_{\omega }
\label{eq2.5}
\end{equation}
and the Eqs. (\ref{eq2.1})-(\ref{eq2.4}), it is possible to obtain
the following general Green's functions
\begin{equation}
\langle\langle A_n;B\rangle\rangle_{\omega }=\sum_m\widetilde{G}_{nm}(\omega)
\langle [ A_{m},B]_{\left( +\right) }\rangle
\label{eq2.6}.
\end{equation}
In the particular case, where $B=A_m^{\dagger}$, the elements of the Green's
function matrix $\bf G$ are given by Eq. (\ref{eq2.5}). Thus, using the matrices
$\bf E$ and $\bf N$, the matrix $\bf G$ is given by
\begin{equation}
{\bf G}\left( \omega \right) =\widetilde{\bf G}(\omega){\bf N}
\label{eq2.7}
\end{equation}
where
\begin{equation}
\widetilde{\bf G}\left( \omega \right) ={\bf N}(\omega {\bf N}-{\bf E})^{-1}
\label{eq2.8}.
\end{equation}

Considering the fact that the operators of the set $\{A_n\}$  describe the particle
excitations of the system, the choice of these operators is very relevant
to study the physical properties of the system.
In order to discuss superconductivity, Beenen and Edwards, in their approach
with the one-band Hubbard model,  mixed electron and hole operators
and evaluated anomalous correlation functions \cite{ref9}. Therefore, using a set of four operators
$\{c_{i\sigma },n_{i-\sigma }c_{i\sigma },c_{i-\sigma }^{\dag},$ $n_{i\sigma }c_{i-\sigma }^{\dag}\}$,
it has been obtained a four-pole approximation to the Green's functions.
However,  in order to discuss the role of the hybridization, it is necessary
to adapt the formalism to include a $p$-operator in the original set of operators used by
Beenen and Edwards. Thus, the new set of operators is given by
\begin{equation}
\left\{d_{i\sigma },n_{i-\sigma }^{d}d_{i\sigma },d_{i-\sigma }^{\dag},
n_{i\sigma }^{d}d_{i-\sigma }^{\dag},p_{i\sigma }\right\}
\label{eq2.9}.
\end{equation}

In the present work, only the singlet pairing is considered, and particularly the
{\it d}-wave symmetry. In this particular case, $\langle d_{i-\sigma }d_{i\sigma }\rangle = 0 $
and $\displaystyle\sum_l\langle d_{i-\sigma }d_{l\sigma }\rangle = 0 $, where $l$
are the nearest neighbors of $i$. Using the set of operators given in Eq. (\ref{eq2.9}),
and introducing the symmetries discussed above, the elements of the energy matrix defined in
Eq. (\ref{eq2.3}) can be obtained as:
{\footnotesize
\begin{eqnarray}
& &{\bf E_5}=\nonumber \\ 
& &\left[
\begin{array}{ccc}
 {\bf E_2}
    &\begin{array}{lr}
      ~0~~~~~~    & 0 \\ \\
      ~0    & {\overline{\gamma}_{k}}
    \end{array}
    &\begin{array}{r}
       V_{k}^{dp} \\ \\
       n_{-\sigma}^dV_{k}^{dp}
    \end{array}\\ \\

    \begin{array}{lr}
      0~~~~~      & 0 \\  \\
      0    & {\overline{\gamma}_{k}}^*
    \end{array}
    &-{\bf E_2}
    &\begin{array}{c}
      0  \\ \\
      0
    \end{array}\\ \\
    \begin{array}{lr}
      V_{k}^{pd} ~~       &  n_{-\sigma}^dV_{k}^{pd}
     \end{array}
     &\begin{array}{lr}
        ~~0~~~~~~~~    & 0
     \end{array}
     &\begin{array}{r}
        ~~~\varepsilon_{p}-\mu + \varepsilon_{k}^{p}
     \end{array}
\end{array}
\right]\nonumber \\
\label{eq2.16}
\end{eqnarray}
}
where $V_{k}^{dp}$ and $V_{k}^{pd}$ are the Fourier transform
of $t_{ij}^{dp}$ and $t_{ij}^{pd}$ respectively. The matrix ${\bf E_2}$
present in the energy matrix ${\bf E_5}$ is given by:
\begin{eqnarray}
{\bf E_2}=\left[
\begin{tabular}{ccc}
$\overline{\varepsilon}_{d} +\varepsilon_{k}^{d} + Un_{-\sigma}^d$ &
$(\overline{\varepsilon}_{d} + \varepsilon_{k}^{d}+U)n_{-\sigma}^d$\\
\\$(\overline{\varepsilon}_{d} + \varepsilon_{k}^{d}+U)n_{-\sigma}^d$ &
$Un_{-\sigma}^d + \Gamma_{k-\sigma}$
\end{tabular}
\right]
\label{eq2.10}
\end{eqnarray}
where $\overline{\varepsilon}_d=\varepsilon_{d}-\mu$.
It is assumed that the system considered here is translationally invariant,
then $n_{-\sigma}^d=n_{i-\sigma}^d$.
The quantity $\Gamma_{k-\sigma}$ is the Fourier transform of
\begin{equation}
\Gamma_{ij-\sigma}=(\varepsilon_{d}-\mu)n_{-\sigma}^d\delta_{ij} + t_{ij}^d
(n_{-\sigma }^d)^2 +  n_{-\sigma}^{d}( 1-n_{-\sigma }^{d})W_{ij-\sigma } .
\label{eq2.11}
\end{equation}
In Eq. (\ref{eq2.11}), the band shift $W_{ij-\sigma }$ is defined as:
\begin{equation}
W_{ij-\sigma }=\frac{t_{ij}^{d}\left( \langle n_{i-\sigma }^{d}n_{j-\sigma
}^{d}\rangle-(n_{-\sigma }^{d})^{2}\right) +\Lambda _{ij\sigma }}{n_{-\sigma
}^{d}( 1-n_{-\sigma }^{d}) }
\label{eq2.12}
\end{equation}
where $\Lambda _{ij\sigma }$ can be separated into two explicit
contributions
\begin{equation}
\Lambda _{ij\sigma }=\Lambda _{ij\sigma }^{d}+\Lambda _{ij\sigma }^{pd}
\label{eq2.13}.
\end{equation}
The terms $\Lambda_{ij\sigma}^{d}$ and $\Lambda_{ij\sigma}^{pd}$ are associated
with the hopping $t_{ij}^{d}$ and the hybridization $t_{ij}^{pd}$, respectively.
The hybridized term of $\Lambda _{ij\sigma }$ may be written as
\begin{equation}
\Lambda _{ij\sigma }^{pd}=\sum_{l}t_{il}^{pd}
[2\langle p_{l-\sigma }^{\dagger}n_{i\sigma}^{d}d_{i-\sigma }
\rangle -\langle p_{l-\sigma }^{\dagger}d_{i-\sigma }
\rangle]\delta _{ij}
\label{eq2.14},
\end{equation}
and the part associated to the hopping $t_{ij}^{d}$ is given by
\begin{eqnarray}
\Lambda _{ij\sigma }^{d}&=& \sum_lt_{il}^d\left \{\langle n_{i\sigma
}^{d}d_{i-\sigma }^{\dagger}d_{l-\sigma }\rangle +\langle n_{i\sigma }^{d}d_{l-\sigma
}^{\dagger}d_{i-\sigma }\rangle\right.\nonumber \\
                        & & +\left.\langle d_{i-\sigma }^{\dagger}
d_{l-\sigma }n_{i-\sigma}^{d}\rangle-\langle d_{l-\sigma }^{\dagger}d_{i-\sigma }
n_{i-\sigma }^{d}\rangle\right\}\delta_{ij}\nonumber \\
                        & & -~t_{ij}^d\{\langle d_{j\sigma }^{\dagger}d_{j-\sigma }^{\dagger}
d_{i-\sigma }d_{i\sigma }\rangle+\langle d_{j\sigma}^{\dagger}d_{i-\sigma }^{\dagger}
d_{j-\sigma }d_{i\sigma }\rangle\}.\nonumber \\
\label{eq2.15}
\end{eqnarray}
The calculation of the correlation functions presented in Eqs. (\ref{eq2.14}) and (\ref{eq2.15})
will be discussed in detail in section \ref{sec:4}.

One of the most important elements of the matrix ${\bf E_5}$  is
$E_{24}={\overline{\gamma}_{k}}$, where
\begin{equation}
\overline{\gamma }_{k}=\sum_{\langle l\rangle i}t_{il}^{d}
e^{i\vec{k}\cdot(\vec{R}_l-\vec{R}_i)}\overline{\gamma }_{il}
\label{eq2.17}
\end{equation}
and
\begin{equation}
\overline{\gamma }_{il}=\langle n_{i-\sigma }^dd_{l\sigma }d_{l-\sigma
}+n_{l\sigma }^dd_{i\sigma }d_{i-\sigma }\rangle
\label{eq2.18}.
\end{equation}
The correlation function $\overline{\gamma }_{k}$ gives the gap of
the superconductor state in the $d-$wave case.

The elements of the normalization matrix ${\bf N_5}$ are given from Eq. (\ref{eq2.4}) as:
\begin{equation}
N_{11}=N_{33}=N_{55}=1
\label{eq2.19}
\end{equation}
and
\begin{equation}
N_{12}=N_{21}=N_{22}=N_{34}=N_{43}=N_{44}=n_{-\sigma }^d.
\label{eq2.20}
\end{equation}
The remaining elements of the normalization matrix ${\bf N_5}$,
due to the {\it d}-wave symmetry and the anticommution rules, have been found
to be zero.

 Using the energy and the normalization matrices ${\bf E_5}$ and ${\bf N_5}$, respectively, the  matrix
 Green's function $\bf G_5$ defined in Eq. (\ref{eq2.7}) can be obtained. For simplicity, only
the \textit{most relevant} elements (for the purposes of this work) of this  $(5\times 5)$
${\bf G_5}$ matrix  are shown. Following the Roth's notation \cite{ref8}, the correlation
function $\langle BA\rangle$ is related to the Green's function
$\langle\langle A;B\rangle\rangle_{\omega }$ as:
\begin{equation}
\langle BA\rangle={\cal F}_{\omega}\langle\langle A;B\rangle\rangle_{\omega }
\equiv\frac{1}{2\pi i}\oint d\omega f(\omega)\langle\langle A;B\rangle\rangle_{\omega }
\label{eq2.21},
\end{equation}
where $f(\omega)$ is the Fermi function. The chemical potential $\mu$ is obtained
in the standard way, using the element $G_{\vec {k}\sigma}^{11}$ of the matrix
${\bf G_5}$ and the relation given in Eq. (\ref{eq2.21}). The matrix element
$G_{\vec {k}\sigma}^{11}$ is given by
\begin{equation}
G_{k\sigma }^{11}(\omega)=\frac{(\omega - E_{55})\left[A\left(\omega\right)-
(\omega + E_{11}){\overline{\gamma}_{k}}^2\right]}
{\overline{D}\left( \omega \right)},
\label{eq2.22}
\end{equation}
where $E_{11}$ and $E_{55}$ are elements of the energy matrix $\bf E_5$, defined in
Eq. (\ref{eq2.16}). In Eq. (\ref{eq2.22}), it is also necessary to introduce the following
definitions:
\begin{eqnarray}
A(\omega)&=&(n_{-\sigma }^d)^2(1-n_{-\sigma }^d)^2\nonumber\\
         & &\times (\omega^3 + \alpha_{k\sigma}^{(1)}\omega^2
+ \alpha_{k\sigma}^{(2)}\omega + \alpha_{k\sigma}^{(3)})
\label{eq2.23}
\end{eqnarray}
with
\begin{equation}
\alpha_{k\sigma}^{(1)}=E_{11}
\label{eq2.24},
\end{equation}
\begin{equation}
\alpha_{k\sigma}^{(2)}={\cal{Z}}_{k\sigma}^{(1)}{\cal{Z}}_{k\sigma}^{(2)}-
({\cal{Z}}_{k\sigma}^{(1)}+{\cal{Z}}_{k\sigma}^{(2)})({\cal{Z}}_{k\sigma}^{(1)}
+{\cal{Z}}_{k\sigma}^{(2)}-E_{11})
\label{eq2.25}
\end{equation}
and
\begin{equation}
\alpha_{k\sigma}^{(3)}=-{\cal{Z}}_{k\sigma}^{(1)}{\cal{Z}}_{k\sigma}^{(2)}
({\cal{Z}}_{k\sigma}^{(1)}+{\cal{Z}}_{k\sigma}^{(2)}-E_{11})
\label{eq2.26}.
\end{equation}
The quantities ${\cal{Z}}_{k\sigma}^{(1)}$ and ${\cal{Z}}_{k\sigma}^{(2)}$ are defined as
\begin{equation}
{\cal{Z}}_{k\sigma}^{(1)}=\frac{U+2(\varepsilon_{d}-\mu)+\varepsilon_{k}^{d}+W_{k-\sigma}}{2}
-\frac{\Delta_{k\sigma}}{2}
\label{eq2.27}
\end{equation}
and,
\begin{equation}
{\cal{Z}}_{k\sigma}^{(2)}={\cal{Z}}_{k\sigma}^{(1)}+\Delta_{k\sigma}
\label{eq2.28}.
\end{equation}
In the particular case, when $\varepsilon_d$, $\overline{\gamma}_{k}$ and $t_{ij}^{pd}$ are zero,
${\cal{Z}}_{k\sigma}^{(1)}$ and ${\cal{Z}}_{k\sigma}^{(2)}$ represent the quasi-particle bands in the
paramagnetic normal state of the one-band Hubbard model. The term $\Delta_{k\sigma}$ is given by:
\begin{equation}
\Delta_{k\sigma}=\sqrt{(U+W_{k-\sigma }-\varepsilon_{k}^{d})^2
+4n_{-\sigma}^dU(\varepsilon_{k}^{d}-W_{k-\sigma })}
\label{eq2.29}
\end{equation}
where $W_{k-\sigma }$ is the Fourier transform of $W_{ij-\sigma }$ given in Eq. (\ref{eq2.12}).
The denominator of the Green's function $G_{\vec{k}\sigma}^{11}$ given in Eq. (\ref{eq2.22}) is
defined as:
\begin{eqnarray}
\overline{D}(\omega)&=&(\omega - E_{55})D(\omega)-
V_k^{dp}V_k^{pd}\left[A(\omega)
-(\omega + E_{11}){\overline{\gamma}_{k}}^2\right]\nonumber \\
\label{eq2.30}
\end{eqnarray}
where
\begin{equation}
D(\omega)={\cal{D}}(\omega)-{\overline{\gamma}_{k}}^2(\omega^2-E_{11}^2)
\label{eq2.31}
\end{equation}
with
\begin{eqnarray}
{\cal{D}}(\omega)&=&[(\omega-E_{11})(\omega n_{-\sigma}^d-E_{22})-(\omega
n_{-\sigma}^d-E_{12})^2]\nonumber\\
& &\times[(\omega+E_{11})(\omega n_{-\sigma}^d+E_{22})-(\omega
n_{-\sigma}^d+E_{12})^2].\nonumber\\
\label{eq2.32}
\end{eqnarray}
In Eq. (\ref{eq2.32}), $E_{12}$ and $E_{22}$ are elements
of the energy matrix $\bf E_2$ given in Eq. (\ref{eq2.10}).
The use of a set of five operators $A_n$ results in a five-pole approximation
to the Green's functions. Then, the $\overline{D}(\omega)$
defined in Eq. (\ref{eq2.30}) may be also written as:
\begin{eqnarray}
\overline{D}\left( \omega \right)&=&(n_{-\sigma }^d)^2(1-n_{-\sigma }^d)^2(\omega-E_{1k})
(\omega-E_{2k})(\omega-E_{3k})\nonumber \\
& &\times(\omega-E_{4k})(\omega-E_{5k})
\label{eq2.33}
\end{eqnarray}
where the quasi-particle bands $E_{pk}$ (with $p=1,..,5$) satisfy
$\overline{D}=det(\omega{\bf{N_5}}-{\bf{E_5}})=0$.
Therefore, the resulting Green's function can be written as a sum of five terms:
\begin{equation}
G_{k\sigma }^{11}(\omega)=\sum_{s=1}^5\frac{Z_{p\vec{k}\sigma}}{\omega-E_{p\vec{k}\sigma}}
\label{eq10.3}
\end{equation}
where $Z_{p\vec{k}\sigma}$ express the spectral weights which satisfy
\begin{equation}
Z_{1\vec{k}\sigma}+Z_{2\vec{k}\sigma}+Z_{3\vec{k}\sigma}+Z_{4\vec{k}\sigma}+Z_{5\vec{k}\sigma}=1
\label{eq10.4}.
\end{equation}
\section{Calculation of the gap function using the factorization procedure}
\label{sec:3}

In the case of $d$-wave symmetry, the traditional correlation function
$\langle d_{i-\sigma }d_{i\sigma }\rangle$ is always zero. Therefore,
this correlation function can not be used to determinate the pairing gap in the $d$-wave channel \cite{ref9,ref14}.
In the factorization procedure proposed by Beenen and Edwards in Ref. \cite{ref9},
the correlation function given by Eq. (\ref{eq2.18}) is rewritten as
\begin{equation}
\overline{\gamma }_{il}=[\langle d_{i-\sigma}^{\dagger}d_{l-\sigma}\rangle
+\langle d_{l\sigma}^{\dagger}d_{i\sigma}\rangle]\langle d_{i-\sigma}d_{l\sigma}\rangle
\label{eq3.0}
\end{equation}
where the symmetry $\overline{\gamma }_{il}=\overline{\gamma }_{li}$ is conserved,
and the products $d_{l\sigma }d_{l-\sigma }$ and $d_{i-\sigma }d_{i\sigma }$ are split up.
It is also introduced
\begin{equation}
n_{01\sigma}^d=\langle d_{i-\sigma}^{\dagger}d_{l-\sigma}\rangle=
\langle d_{l\sigma}^{\dagger}d_{i\sigma}\rangle
\label{eq3.1}
\end{equation}
which allows to rewrite Eq. (\ref{eq3.0}) as
\begin{equation}
\overline{\gamma }_{il}=2n_{01\sigma}^d\langle d_{i-\sigma}d_{l\sigma}\rangle
\label{eq3.2}
\end{equation}
where $n_{01\sigma}^d$ can be calculated from $G_{k\sigma }^{11}$.

Considering the {\it d}-wave symmetry, the Fourier transform of $\overline{\gamma }_{il}$
given by Eq. (\ref{eq2.17}) becomes
\begin{equation}
\overline{\gamma }_{k}=\overline{g}\left[\cos{(k_xa)}-\cos{(k_ya)}\right]
\label{eq3.3}
\end{equation}
where
\begin{equation}
\overline{g}=2t^{d}\overline{\gamma}
\label{eq3.4}
\end{equation}
is the gap-function amplitude. Due to the $d$-wave symmetry, $\overline{\gamma }_{il}=+\overline{\gamma }$
for $\vec{R}_i-\vec{R}_l$ in the $x$ direction and $\overline{\gamma }_{il}=-\overline{\gamma }$
when $\vec{R}_i-\vec{R}_l$ is in the $y$ direction. The Fourier
transform of the correlation function $\langle d_{i-\sigma }d_{l\sigma }\rangle$
is given by
\begin{equation}
\langle d_{i-\sigma}d_{l\sigma}\rangle=\frac{1}{L}\sum_{k}
e^{i\vec{k}\cdot(\vec{R}_l-\vec{R}_i)}\langle d_{k-\sigma}d_{k\sigma}\rangle
\label{eq3.5}
\end{equation}
where $L$ is the number of sites in the system. The correlation function
$\langle d_{k-\sigma}d_{k\sigma}\rangle $ can be evaluated using the Green's function
$G_{k\sigma}^{13}$ and the relation given by Eq. (\ref{eq2.21}).
The Green's function $G_{k\sigma}^{13}$ can be rewritten as:
\begin{equation}
G_{k\sigma }^{13}(\omega)=-\overline{\gamma }_{k}U^2F_{k\sigma }^{13}
\label{eq3.6}
\end{equation}
where
\begin{equation}
F_{k\sigma }^{13}(\omega)=\frac{(n_{-\sigma }^d)^2(1-n_{-\sigma }^d)^2(\omega - E_{55})}
{\overline{D}\left( \omega \right)}
\label{eq3.7}
\end{equation}
and $\overline{D}(\omega)$ is defined in Eq. (\ref{eq2.30}).

Combining the equation (\ref{eq2.17}) with the Eqs. (\ref{eq3.2}) to (\ref{eq3.6}),
the gap equation can be written as:
\begin{equation}
\overline{\gamma}_{k}=-\overline{\gamma}_{k}2n_{01\sigma}^dt^{d}U^2I_{\sigma}
\label{eq3.8}
\end{equation}
where
\begin{equation}
I_{\sigma}=\frac{1}{2\pi i}{\displaystyle \oint} f(\omega)
F_{\sigma }(\omega)d\omega              
\label{eq3.81}
\end{equation}
with
\begin{equation}
F_{\sigma}(\omega)=\frac{1}{L}
\sum_{\vec{q}}\left[\cos{(\vec{q}_xa)}-\cos{(\vec{q}_ya)} \right]^2
F_{q\sigma }^{13}(\omega)
\label{eq3.9}.
\end{equation}

\section{Definition and calculation of the band shifts}
\label{sec:4}

Using the definition (\ref{eq2.13}) in Eq. (\ref{eq2.12}), the band shift $W _{ij\sigma }$
can be written as:
\begin{equation}
W _{ij-\sigma }=W_{ij-\sigma }^{d}+W_{ij-\sigma }^{pd}
\label{eq4.0}
\end{equation}
where
\begin{equation}
W_{ij-\sigma }^d=\frac{t_{ij}^{d}[ \langle n_{i-\sigma }^{d}n_{j-\sigma
}^{d}\rangle-(n_{-\sigma }^{d})^{2}] +\Lambda _{ij\sigma }^d}{n_{-\sigma
}^{d}( 1-n_{-\sigma }^{d}) }
\label{eq4.1}
\end{equation}
and
\begin{equation}
W_{ij-\sigma }^{pd}=\frac{\Lambda _{ij\sigma }^{pd}}{n_{-\sigma
}^{d}( 1-n_{-\sigma }^{d}) }
\label{eq4.2}.
\end{equation}
The quantity $\Lambda_{ij\sigma }^{pd}$ is given by Eq. (\ref{eq2.14}).
The correlation function $\langle p_{l-\sigma }^{\dagger}d_{i-\sigma }
\rangle$ present in $\Lambda_{ij\sigma }^{pd}$ can be obtained from de
Green's function
\begin{equation}
G_{k\sigma }^{15}(\omega)=\frac{\left[A\left(\omega\right)-
(\omega + E_{11}){\overline{\gamma}_{k}}^2\right]V_k^{pd}}
{\overline{D}\left( \omega \right)}
\label{eq4.3}.
\end{equation}
The remaining correlation function $\langle p_{l-\sigma }^{\dagger}n_{i\sigma}^{d}d_{i-\sigma }
\rangle$ present in $\Lambda_{ij\sigma }^{pd}$ is calculated from the Green's function
\begin{equation}
G_{k\sigma }^{25}(\omega)=\frac{n_{-\sigma }^d\left[B\left(\omega\right)-
(\omega + E_{11}){\overline{\gamma}_{k}}^2\right]V_k^{dp}}
{\overline{D}\left( \omega \right)}
\label{eq4.4},
\end{equation}
where
\begin{equation}
B\left(\omega\right)=A\left(\omega\right)
+ n_{-\sigma }^d(1-n_{-\sigma }^d)^2U{\cal{D}}_1(\omega)
\label{eq4.5}
\end{equation}
with $A\left(\omega\right)$ defined in Eq. (\ref{eq2.23}). The quantity ${\cal{D}}_1(\omega)$, in terms
of the elements of the energy matrix (\ref{eq2.16}), is given by:
\begin{equation}
{\cal{D}}_1(\omega)=(\omega-E_{11})(\omega n_{-\sigma}^d-E_{22})-(\omega
n_{-\sigma}^d-E_{12})^2
\label{eq4.6}.
\end{equation}

The Green's function $G_{k\sigma }^{25}$ tends to zero as $U\rightarrow \infty$, consequently,
the correlation function $\langle p_{l-\sigma }^{\dagger}n_{i\sigma}^{d}d_{i-\sigma }\rangle$
also vanishes recovering the result of Ref. \cite{ref6} for $\Lambda _{ij\sigma }^{pd}$.

The quantity $\Lambda_{ij\sigma}^d$ present in Eq. (\ref{eq4.1}) is given
by Eq. (\ref{eq2.15}).
The Fourier transform of $W_{ij\sigma}^d$ is given by:
\begin{equation}
W_{k\sigma}^d=\sum_{\langle j \rangle i}
e^{i\vec{k}\cdot(\vec{R}_j-\vec{R}_i)}W_{ij\sigma}^d
\label{eq4.7}.
\end{equation}
Substituting Eq. (\ref{eq2.15}) into Eq. (\ref{eq4.1}) and then
putting the result into Eq. (\ref{eq4.7}), the Fourier transform
of $W_{ij-\sigma}^d$ can be written as:
\begin{eqnarray}
W_{k\sigma}^d&=&-\frac{1}{n_{\sigma}^d(1-n_{\sigma}^d)}\sum_{j\neq 0}t_{0j}^{d}
\langle d_{0\sigma }^{\dagger}d_{j\sigma}(1-n_{0-\sigma }^{d}-n_{j-\sigma }^{d})\rangle\nonumber \\
& &+\sum_{j\neq 0}t_{0j}^{d}e^{i\vec{k}\cdot\vec{R}_j}\left \{\langle n_{j\sigma }^{d}n_{0\sigma}^{d}
\rangle -\langle n_{0\sigma }^{d}\rangle^{2}\right. \nonumber \\
& &\left. +~\langle d_{j\sigma }^{\dagger}d_{j-\sigma }
d_{0-\sigma }^{\dagger}d_{0\sigma }\rangle - \langle d_{j\sigma}^{\dagger}d_{j-\sigma }^{\dagger}
d_{0-\sigma }d_{0\sigma }\rangle \right\}
\label{eq4.8}.
\end{eqnarray}

The correlation functions present in $W_{k\sigma}^d$ are evaluated
following the original Roth's procedure \cite{ref8}. Introducing
extra operators $B_{i\sigma}$, the correlation functions of the
form $\langle A_nB_{i\sigma} \rangle$ can be calculated by using
Eqs. (\ref{eq2.6}) and (\ref{eq2.21}). In Refs.
\cite{ref9,ref11}, the sum present in Eq. (\ref{eq2.6}) has
been considered only over the operators which describe the normal
state of the system. In the present work, the sum includes also
the hole operators which describe the superconducting properties
of the system. Thus, $W_{k\sigma}^d$ is given by:
%
%
\begin{equation}
n_{\sigma}^d(1-n_{\sigma}^d)W_{\vec{k}\sigma }^d=h_{1\sigma}
+\sum_{j\neq 0}t_{0j}^{d}e^{i\vec{k}\cdot\vec{R}_j}(h_{2j\sigma}
+h_{3j\sigma})
\label{eq4.9}
\end{equation}
where the term $h_{3j\sigma}$ is directly related to the gap function $\overline{\gamma}_{\vec{k}}$
through the Green's functions $G_{\vec{k}\sigma}^{13}$ and $G_{\vec{k}\sigma}^{14}$ (see Appendix A).
The quantities $h_{1\sigma} $, $h_{2j\sigma}$ and $h_{3j\sigma}$ are given in Appendix A.

\section{Results}
\label{sec:5}

In this section, the numerical results obtained in this work are presented.
One of the most important parameters of the model given in Eq. (\ref{eq2.0})
is the $d-p$ hybridization \cite{ref11,ref15}, which is defined as
\begin{equation}
V_{\vec{k}}^{dp}=-iV_0^{dp}[\sin(k_xa)-\sin(k_ya)]
\label{eq5.0}.
\end{equation}
In this work, as in Ref. \cite{ref15}, the hybridization has been assumed $\vec{k}$-independent
$(V_0^{dp})^2\equiv \langle V_{\vec{k}}^{dp} V_{\vec{k}}^{pd} \rangle $, where $\langle ...\rangle$
is the average over the Brillouin zone. In Ref. \cite{ref16} Sengupta and Ghatak have also used
a $\vec{k}$-independent hybridization due to the fact that the pairs occur within a small energy
interval around the Fermi level, therefore the dispersion of the hybridization can be neglected.

The total occupation number is given by $n_T=n_{\sigma}^d+n_{-\sigma}^d$, where $n_{\sigma}^d$
is obtained combining $G_{k\sigma }^{11}$ (Eq. (\ref{eq2.22})) and the relation given
in Eq. (\ref{eq2.21}). The charge transfer energy
$\Delta=\varepsilon_p - \varepsilon_d$ is positive. This means that the first hole added to the
system will energetically prefer to occupy the $d$-orbital of the copper ions \cite{ref4}.
All results presented in this section are obtained with $\varepsilon_d =0$ and
$\varepsilon_p =3.6 eV$. Consequently, $\Delta= \nolinebreak 3.6eV$, as estimated in Ref. \cite{ref17}.

\begin{figure}[t]
\resizebox{.5\textwidth}{!}{
\includegraphics[angle=270,width=7.5cm]{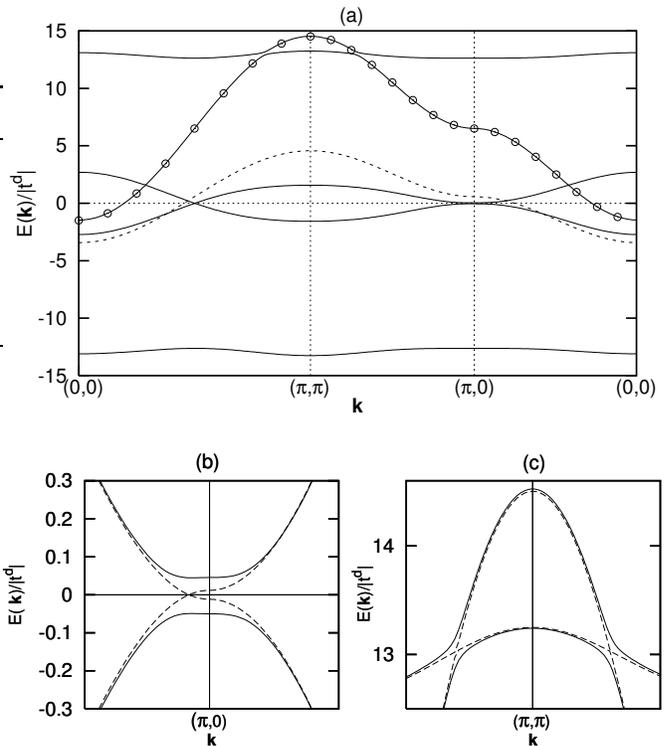}}
\caption{(a) The quasi-particle bands for $U=12|t^d|$, $V_0^{pd}=0.2|t^d|$ and $n_T=0.76$.
(b) The electron and hole bands in the neighborhood of the $(\pi,0)$ point where
the gap structure is relevant. The dashed lines show the result for $V_0^{pd}=0.0$
while the solid {bf lines} correspond to $V_0^{pd}=0.2|t^d|$. (c) shows the hybridization ($V_0^{pd}$) gap
near the point $(\pi,\pi)$.}
\label{fig1}
\end{figure}
\begin{table}
\caption{The different approaches considered to evaluate the band  shift $W_{\vec{k}\sigma}$.}
\label{tab:1}       
\begin{tabular}{llllll}
\hline\noalign{\smallskip}
& $U$ & $\overline{\gamma}_{\vec{k}}$ & $T$ & $V_0^{pd}$ & $h_{3j\sigma}$ \\
\noalign{\smallskip}\hline\noalign{\smallskip}
Beenen and\\
Edwards \cite{ref9} & $finite$ & 0 &  0 &  0 &  0 \\ \\
Ref.\cite{ref11}& $\infty$ & 0 & 0 & $finite$ & 0 \\ \\
Present work & $finite$ & $finite$ & $finite$ & $finite$ & $finite$ \\
\noalign{\smallskip}\hline
\end{tabular}
\end{table}
\begin{figure}
\resizebox{.5\textwidth}{!}{
\includegraphics[angle=270,width=7.5cm]{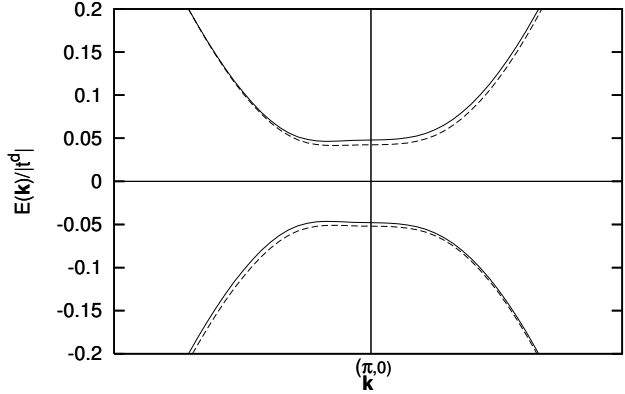}}
\caption{The electron and hole bands in the region close to the $(\pi,0)$ point
for $U=12|t^d|$ and $n_T=0.76$. The solid lines correspond to $V_0^{pd}=0.0$,
while the dashed lines show the result for $V_0^{pd}=0.3|t^d|$.}
\label{fig1.1}
\end{figure}

As discussed in Refs. \cite{ref9,ref8}, the band shift
$W_{\vec{k}\sigma}$ (see Eq. (\ref{eq4.0})) can be evaluated considering different
approximations. In the limit $U\rightarrow \infty$, some terms of
the band shift vanish (see Ref. \cite{ref8}). In
Ref. \cite{ref11}, the present authors estimate
$W_{\vec{k}\sigma}$ in the limit of $U\rightarrow \infty$, but
with finite $U$ in other parts of the problem. In Ref.
\cite{ref9}, Beenen and Edwards evaluated $W_{\vec{k}\sigma}$
in the normal state (where $\overline{\gamma}_{\vec{k}}=0$) and
considering $T$ equal to zero and finite $U$ using the one-band Hubbard model (hybridization null).
In the present work, the
correlation functions present in $W_{\vec{k}\sigma}$ given in Eq.
(\ref{eq4.8}) are evaluated following closely the procedure used
by Roth in Ref. \cite{ref8}. Nevertheless, here, also the
hole operators given in the set of Eq. (\ref{eq2.9}) are used to
evaluate the correlation functions. As consequence, a new term  ($h_{3j\sigma}$)
appears in $W_{\vec{k}\sigma}$ (see Eq. (\ref{eq4.9})). The approximations used to evaluate
$W_{\vec{k}\sigma}$ are shown in Table \nolinebreak \ref{tab:1}.
In figure \ref{fig1}(a), the quasi-particle bands $E_{pk}$, with $p=1..5$ (see Eq. (\ref{eq2.33})),
are plotted along the symmetry lines $(0,0)-(\pi,\pi)-(\pi,0)-(0,0)$,
in the two-dimensional Brillouin zone. The quasi-particle energies $E_{pk}$,
in the superconducting state, are relative to the chemical potential $\mu$.
The circles {bf show} the $\varepsilon_{\vec{k}}^p$ band, where the center of
$\varepsilon_{\vec{k}}^p$ is shifted by $\varepsilon_p = 3.6 eV$ relative to the zero of energy.
All results shown in this paper are obtained with  $t^p=2t^d$.
The dashed line corresponds to the noninteracting $(U=0)$
band $\varepsilon_{\vec{k}}^d$ relative to the noninteracting chemical potential.
The figure \ref{fig1}(b) shows the superconducting gap between the electron and hole bands
in the neighborhood of the $(\pi,0)$ point, while on the $k_x=k_y$ diagonal (Fig. \ref{fig1}(a))
the gap is zero. This fact reflects the $d$-wave symmetry proposed in this work. The dashed lines
show the absence of the gap in the normal state. In figure \ref{fig1}(c),
the region near to the $(\pi,\pi)$ point shows the gaps produced by the $d-p$ hybridization $V_0^{pd}$.
The dashed lines show the result for $V_0^{pd}=0$.
In figure  \ref{fig1.1}, the electron and hole quasi-particle bands are shown for two
different values of hybridization. As can be observed,
the hybridization shifts the quasi-particle bands to lower energy by  breaking the symmetry
{in relation to the} $\vec{k}$ axis.

\begin{figure}[ht]
\resizebox{.4\textwidth}{!}{
\includegraphics[angle=270,width=7.5cm]{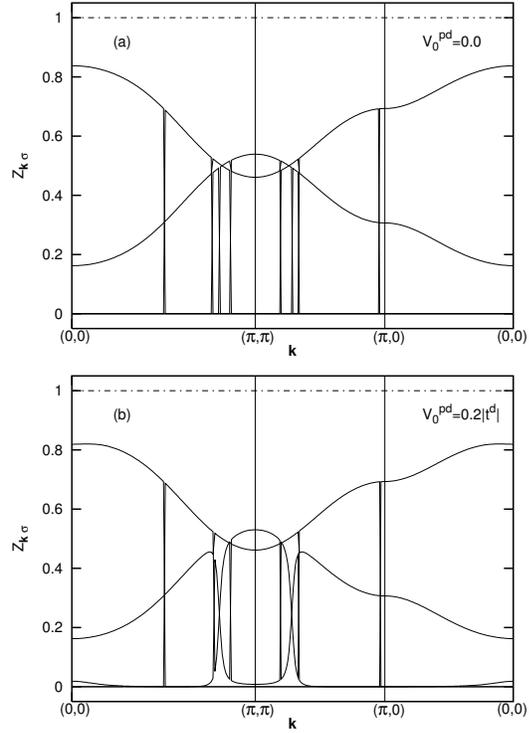}}
\caption{The spectral weights $Z_{p\vec{k}\sigma}$ for $U=12|t^d|$, $n_T=0.76$, $T=0$ and
two different values of hybridization.}
\label{fig2}
\end{figure}

The figure \ref{fig2} shows the spectral weights  $Z_{p\vec{k}\sigma}$ for two different hybridizations.
The dashed line corresponds to the sum of the five spectral weights which is equal to
one  (see Eq. (\ref{eq10.4})).
In figure  \ref{fig2}(b), the effects of the hybridization on the spectral weights are shown. Such
effects cause a small change in the chemical potential and consequently in the superconductivity.

\begin{figure}[ht]
\resizebox{.5\textwidth}{!}{
\includegraphics[angle=270,width=7.5cm]{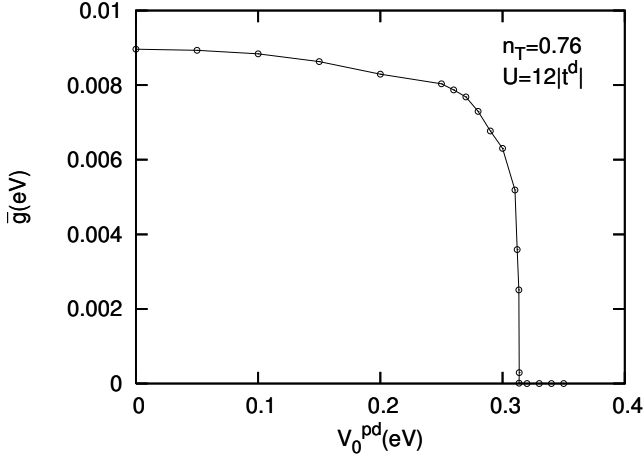}}
\caption{Behavior of the gap function amplitude $\overline{g}$ as
a function of the hybridization $V_0^{pd}$ for T=0.004eV.}
\label{fig3}
\end{figure}

The figure \ref{fig3} shows the behavior of gap
function amplitude $\overline{g}$ as a function of the
hybridization $V_0^{pd}$. It is clear that there is a decreasing
of $\overline{g}$ with increasing $V_0^{pd}$.
\begin{figure}[ht]
\resizebox{.5\textwidth}{!}{
\includegraphics[angle=270,width=7.5cm]{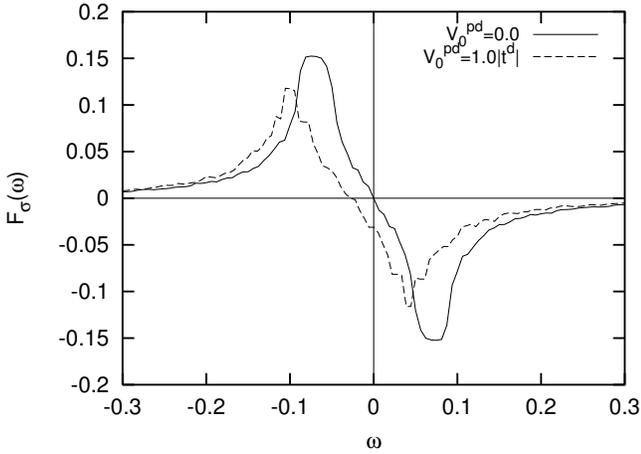}}
\caption{Function $F_{\sigma}(\omega)$ (defined in Eq. (\ref{eq3.9})) for $U=12|t^d|$, $n_T=0.76$,
$T=0$ and two different hybridizations.}
\label{fig4}
\end{figure}

The analysis of the function $F_{\sigma}(\omega)$ introduced in
Eq. (\ref{eq3.81}) and defined in Eq. (\ref{eq3.9}) is important
to understand the behavior of the gap function amplitude showed in
figure \ref{fig3}. The figure \ref{fig4} shows the function
$F_{\sigma}(\omega)$ for $T=0$ and two different values of
hybridization. As can be seen in the dashed line, the magnitude of
the function $F_{\sigma}(\omega)$ decreases when the hybridization
is enhanced. Moreover, the function is shifted to lower energy,
breaking the symmetry respect to $\omega =0$.
The symmetry break has been also observed in figure \ref{fig1.1}
for the electron and hole quasi-particle bands.
For $T=0$, the
product $f(\omega)F_{\sigma}(\omega)$ given in Eq. (\ref{eq3.81})
vanishes when $\omega > 0$. That is because the Fermi function
$f(\omega)$ is zero for that range of $\omega$. As consequence of
the shift and the suppression of $F_{\sigma}(\omega)$, the value
of $I_{\sigma}$, which is given by the integral in Eq. (\ref{eq3.81}),
decreases when the hybridization increase. However,
from Eq. (\ref{eq3.8}), it is necessary a minimum value for
$I_{\sigma}$ to obtain a nonzero solution for $\overline{\gamma}$.
But, for very strong values of hybridization, the minimum value
for $I_{\sigma}$ is not reached and only the zero solution exists.

According to this analysis, there is a critical value of
hybridization ($V_{0c}^{pd}$), above which, the
superconductivity is suppressed.
Similar results, which show a critical value for the hybridization,
were also obtained in Ref. \cite{ref18}, for a $\vec{k}$-dependent
hybridization and using the Hartree-Fock approximation for the 
electron-electron interaction.
In Refs. \cite{ref13,ref18}, although the high
$T_c$ was not considered, the hybridization effects play an
important role for resonant states. The discussion above is also valid
if the values of the temperature $T$ are raised with
$V_{0}^{pd}$ constant. The only difference is that in this case
the Fermi function becomes sloping smoothly, changing the product
$f(\omega)F_{\sigma}(\omega)$. The effect of the temperature in
the Fermi function causes a decreasing  of $I_{\sigma}$ and
consequently of $T_c$.
\begin{figure}[ht]
\resizebox{.50\textwidth}{!}{
\includegraphics[angle=270,width=7.5cm]{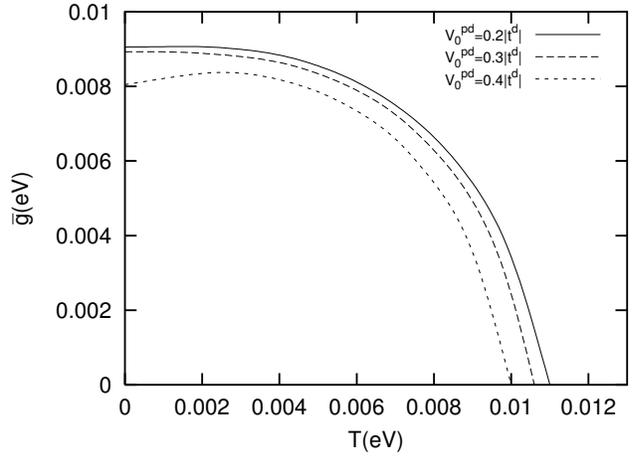}}
\caption{The gap function amplitude $\overline{g}$, as a function of
temperature $T$, for $U=12|t^d|$, $n_T=0.76$ and several values of hybridization.
($t^d=-0.5eV$.)}\label{fig5}
\end{figure}

\begin{figure}[h]
\resizebox{.50\textwidth}{!}{
\includegraphics[angle=270,totalheight=8.0cm]{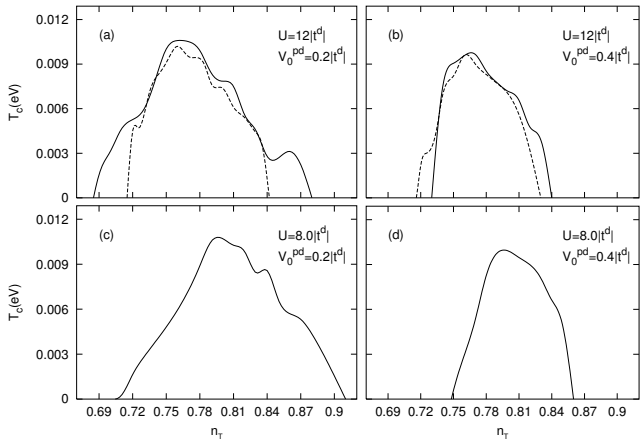}}
\caption{$T_{c}$ as a function of the total occupation number $n_T$.
              In (a) and (b), the dotted
              lines show the previous results from Ref. \cite{ref11} for $U=12|t^d|$.
              The solid lines show the behavior of $T_{c}$ in the present approach.
              The figures (c) and (d) show the present results for $U=8|t^d|$.
              ($t^d=-0.5eV$.)}
\label{fig6}
\end{figure}

Since the hybridization is directly related to the applied
pressure \cite{ref13}, the transition temperature $T_c$ may have a
dependence on pressure through the hybridization. However, the
pressure dependence of $T_c$ is very complicated in high
temperature superconductors. As discussed in Ref. \cite{ref13}, at
least, in conventional superconductivity where the electron
pairing is mediated by phonons, two effects are responsible for the
pressure dependence of $T_c$. The first one is related to the
lattice vibrations, while the second one comes from the electronic
contribution. As long the pressure is increased,
the lattice vibrations tend to increase $T_c$, whereas the effects
of the electronic contribution associated with the hybridization
cause a decreases of $T_c$. The figure \ref{fig5} shows the
function amplitude $\overline{g}$, as a function of temperature
$T$ for several values of hybridization. If the hybridization is
enhanced, the critical temperature $T_c$ decreases. Therefore, the
present results agree with the discussion above in the scenario
where the electronic effects dominate. This behavior for $T_c$ is
also shown in figure \ref{fig6}, in phase diagrams displaying
$T_c$ versus the total occupation number $n_T$. The numerical
results obtained show that there is a critical value of
hybridization where the superconducting phase vanishes. These
results agree with the ones obtained by the authors of Ref.
\cite{ref19} for heavy-fermion superconductivity with an $X$-boson
treatment.

The solid lines in figures \ref{fig6}(a)-(b) show the present
result for $T_c$, where the effects of the temperature, the
superconductivity and the Coulomb interaction have been included in
the calculation of the band shift $W_{\vec{k}\sigma}$. The dashed
lines correspond to the results obtained in Ref. \cite{ref11},
where the band shift has been evaluated considering $T=0$,
$\overline{\gamma}_{\vec{k}}=0$ and $U\rightarrow \infty$ (see
table \ref{tab:1}). The difference between the results can be
explained by the analysis of the Eq. (\ref{eq2.14}) where some of the
correlation functions present in the band shift vanish in the
$U\rightarrow\infty$ limit. In Eq. (\ref{eq2.14}), the correlation
function $\langle p_{l-\sigma
}^{\dagger}n_{i\sigma}^{d}d_{i-\sigma }\rangle$, which is directly
related to the hybridization effects in the band shift, vanishes
for $U\rightarrow\infty$. It is important to highlight that the
correlation functions in Eq. (\ref{eq2.14}) are both negative.
Therefore, for large $U$, the correlation function $\langle
p_{l-\sigma }^{\dagger}n_{i\sigma}^{d}d_{i-\sigma }\rangle$
decreases and the hybridized shift $W_{\sigma}^{pd}$ is enhanced.
However, for intermediate values of $U$, both correlation functions
remain finite. As consequence, the hybridization effects in the
band shift and, therefore, in the superconductivity, are weakened.
The figures \ref{fig6}(c)-(d) show the present results when the
value of $U$ is increased.  The main consequence is, within the
factorization procedure, to shift the window of doping where 
superconductivity is found, as in Ref \cite{ref9}.

\begin{figure}[ht]
\resizebox{.5\textwidth}{!}{
\includegraphics[angle=270,width=7.5cm]{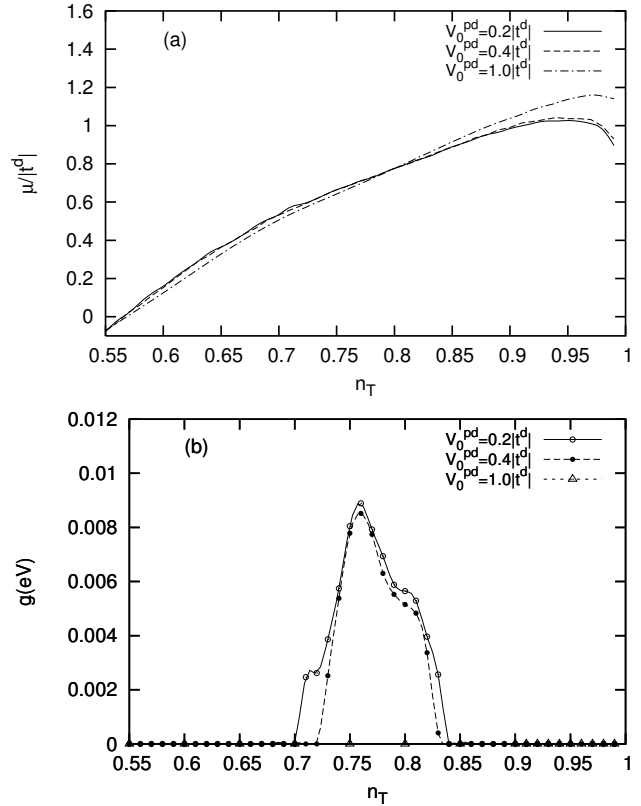}}
\caption{(a) The chemical potential as a function of the total
occupation number for $U=12|t^d|$, $kT=0.004eV$ and different values of hybridization.
(b) The gap function amplitude, as a function of the total
occupation number, with $U$ and $kT$ identical to (a).}
\label{fig7}
\end{figure}

In figure \ref{fig7}(a), the chemical potential is show as a
function of the total occupation number $n_T$ for $U=12|t^d|$ and
two different hybridizations .
In Ref. \cite{ref16}, the authors criticized
the Roth's method because the compressibility $k=\frac{\partial n_T}{\partial\mu}$ is
negative in the vicinity
of half-filling in the Beenen and Edwards result. In Ref. \cite{ref14}, by using a 
composite operator approach and imposing the Pauli principle, the authors have showed 
that the compressibility remains negative. However, they also showed that the pairing 
decreases the strength of the negative compressibility.

In the present work, a careful study about the nature of the negative compressibility
and the effect of the hybridization near half-filling in
Roth's approximation has been carried out. It has been verified that the most important contribution 
to provide negative compressibility  comes from the spin-term $\langle S_jS_i\rangle$ present 
in the $d$-part $W_{\vec{k}\sigma}^d$ of the Roth's band shift $W_{\vec{k}\sigma}$ (see Eqs. \ref{eq4.8} 
and  \ref{Apx26}). In reference \cite{ref9}, it has been showed that the correlation 
function $\langle S_jS_i\rangle$ plays an important role on the flattening of
the quasi-particle bands. The correlation function $\langle S_jS_i\rangle$ increases with occupation and
its effect is pronounced near half-filling. Nevertheless, when the hybridization is present,
the numerical results show that it acts in the sense of suppressing the negative compressibility
near half-filling. Because the hybridization considered here is $\vec{k}$-independent 
\cite{ref15}, the hybridization term $W_{\sigma}^{pd}$ of the band shift is constant within
the Brillouin zone. Its main effect is to shift the poles of the Green's functions and consequently
to change the value of the chemical potential suppressing the negative compressibility. 
In figure \ref{fig7}(a), it is clear that the effect of the hybridization in the chemical 
potential decreases the negative compressibility. 

The figure \ref{fig7}(b) shows the gap function amplitude
$\overline{g}$ as a function of the total occupation number. 
This result agrees with those obtained in figure
\ref{fig3}, where $\overline{g}$ decreases with increasing of
$V_{0}^{pd}$.

\section{Conclusions}
\label{sec:6}

In this work, the Roth's two-pole approximation is extended to study
the superconducting properties of the extended Hubbard model given in Eq. (\ref{eq2.0}).
The quality of the Roth's two-pole approximation had been investigated in a previous work
by Beenen and Edwards \cite{ref9}. In their work, they showed the remarkable agreement between 
the Roth's and the Monte Carlo results \cite{ref6,ref7} for the one-band Hubbard model in
the paramagnetic normal state.  Moreover, the flat bands obtained with Roth's procedure
show a qualitative agreement with the ARPES experiment data \cite{ref1} in cuprates. 
It is important to point out that the flattening observed in the quasi-particle bands which 
produces a peak in the density of states, can be connected with the Van Hove scenario. In cuprate
systems the Van Hove singularity is present in the vicinity of the Fermi energy. Therefore, it is 
believed that the Van Hove scenario play a fundamental role in order to clarify
the mechanism which drives the transition to superconductivity in these interesting materials 
\cite{ref20}. 

The accuracy
of the Roth's results is very related to the adequate evaluation of the band shift. Therefore,
the focus of the present work has been to evaluate the Roth's band shift taking into account relevant 
effects as Coulomb interaction, temperature, superconductivity and hybridization.   
Also, the effect of the hybridization in the superconducting of the model has been studied. This work
has been carried out following the factorization procedure proposed by Beenen 
and Edwards \cite{ref9}. In order to study superconductivity, Beenen and Edwards proposed to 
include hole operators in the original set of operators that describes
the normal state of the system.  
These operators can introduce the pairing formation in the $d$-band.
The factorization procedure proposed by Beenen and Edwards \cite{ref9} and the $d$-wave
symmetry are considered to obtain the gap function amplitude. The hybridization effects
are considered by also including a $p$-operator.  Thus, the set of operators
is enlarged to five, which results in a five-pole approximation to the Green's functions.

The  hybridization effects  present in the  band shift come from
some correlation functions. The important point is that part of
them vanish when $U\rightarrow\infty$, as it have been done in
Ref. \cite{ref11}. In order to consider properly the hybridization
effects, the band shift should be obtained for finite $U$. In
fact, the obtained phase diagrams show that the presence of
superconducting order exists in a larger range of doping when
compared with the $U\rightarrow\infty$ limit \cite{ref11}, for the
same hybridization. Therefore, this result suggests that, in the
$U\rightarrow\infty$ limit, the hybridization effects are
overestimated. That is the ultimate justification for the use of
the factorization procedure \cite{ref9}, which is valid for
intermediated values of $U$ for the gap function.

The Beenen and Edwards's \cite{ref9} results are recovered taking
$V_0^{pd}=0$ in the present work. The hybridization
$V_0^{pd}$  breaks the symmetry between the electron and hole
quasi-particle bands, respect to $\vec{k}$ axis. Also, the gap
amplitude function $\overline{g}$ and the critical temperature $T_c$ are
suppressed with increasing the hybridization $V_0^{pd}$.
The results show that the chemical potential does not change
significantly away the half-filling. However, near half-filling, it is showed 
that the negative compressibility decreases with increasing $V_0^{pd}$. The correlation
functions present in the $d$-part of the band shift
$W_{\vec{k}\sigma}$ were discussed in detail. When the hole
operators are also considered to obtain this correlation functions, a
new term appears in the $d$-part of the band shift
$W_{\vec{k}\sigma}$. The new term is directly associated with the
superconducting properties of the system. Nevertheless, this term
is quite small and therefore may be disregarding in the
calculation of the band shift.

\subsection*{Acknowledgments}

The authors are grateful to the Grupo de F\'{\i}sica Estat\'{\i}stica-IFM,
Universidade Federal de Pelotas where part of numerical calculations were performed. This
work was partially supported by the Brazilian agencies CAPES (Coordena\c{c}\~ao de
Aperfei\c{c}oamento de Pessoal de N\'{\i}vel Superior) and FAPERGS (Funda\c{c}\~ao
de Amparo \`a Pesquisa do Rio Grande do Sul).

\renewcommand{\thesection}{}
\renewcommand{\theequation}{\Alph{section}.\arabic{equation}}
\setcounter{equation}{0}
\setcounter{section}{0}

\appendix

\section{Appendix}
The correlation functions present in the band shift $W_{\vec{k}\sigma}^d$
can be evaluated by introducing extra $B$ operators, as in the original Roth's procedure.
Combining the Eq. (\ref{eq2.6}) and the relation given in Eq. (\ref{eq2.21}), it is possible
to write
\begin{equation}
\langle BA_n \rangle = {\cal{F}}_{\omega}\sum_m \widetilde{G}_{nm}(\omega) \langle [A_m,B]_{(+)}\rangle
\label{Apx1},
\end{equation}
where $A_n$ and $A_m$ are members of the set of operators given in Eq. (\ref{eq2.9}).
For evaluate $\langle n_{j\sigma }^{d}n_{0\sigma}^{d}\rangle - (n_{0\sigma}^{d})^2$,
it has been necessary to introduce the following $B$ operators:
\begin{equation}
B_{\vec{k}j\sigma}^{(1)}=\frac{1}{\sqrt{L}}\sum_ie^{-i\vec{k}\cdot\vec{R}_i}n_{i+j\sigma}^{d}d_{i\sigma
}^{\dagger} \label{Apx2}
\end{equation}
and
\begin{equation}
B_{\vec{k}j\sigma}^{(2)}=\frac{1}{\sqrt{L}}\sum_ie^{-i\vec{k}\cdot\vec{R}_i}n_{i+j-\sigma}^{d}d_{i\sigma
}^{\dagger} \label{Apx3}.
\end{equation}

By considering the operator given in the Eq. (\ref{Apx2}), the correlation function
$\langle n_{j\sigma }^{d}n_{0\sigma}^{d}\rangle$ can be written as:
\begin{equation}
\langle n_{j\sigma }^{d}n_{0\sigma}^{d}\rangle=
\frac{1}{L}\sum_{\vec k}\langle
B_{\vec{k}j\sigma}^{(1)}d_{\vec{k}\sigma}\rangle \label{Apx4}
\end{equation}
where the right side of Eq. (\ref{Apx4}) may be obtained using the
relation given by Eq. (\ref{Apx1}). Therefore, it is necessary to
evaluate the anticommutators $[A_{m},B_{\vec{k}l\sigma}^{(1)}]_{(+)}$ for the set of operators
$A_{m}$ given in Eq. (\ref{eq2.9}). For $m=1..5$, the $A$ operators are given by:
\begin{equation}
A_{1\vec{k}\sigma}=\frac{1}{\sqrt{L}}\sum_le^{i\vec{k}\cdot\vec{R}_l}d_{l\sigma}
\label{Apx5},
\end{equation}
\begin{equation}
A_{2\vec{k}\sigma}=\frac{1}{\sqrt{L}}\sum_le^{i\vec{k}\cdot\vec{R}_l}n_{l-\sigma}^dd_{l\sigma}
\label{Apx6},
\end{equation}
\begin{equation}
A_{3\vec{k}\sigma}=\frac{1}{\sqrt{L}}\sum_le^{i\vec{k}\cdot\vec{R}_l}d_{l-\sigma}^{\dagger}
\label{Apx7},
\end{equation}
\begin{equation}
A_{4\vec{k}\sigma}=\frac{1}{\sqrt{L}}\sum_le^{i\vec{k}\cdot\vec{R}_l}n_{l\sigma}^dd_{l-\sigma}^{\dagger}
\label{Apx8}
\end{equation}
and
\begin{equation}
A_{5\vec{k}\sigma}=\frac{1}{\sqrt{L}}\sum_le^{i\vec{k}\cdot\vec{R}_l}p_{l\sigma}
\label{Apx9}.
\end{equation}
Thus, the following results have been obtained
\begin{equation}
\langle
[A_{1\vec{k}\sigma},B_{\vec{k}j\sigma}^{(1)}]_{(+)}\rangle=n_{0\sigma}^{d}
-e^{i\vec{k}\cdot \vec{R}_j}\langle
d_{0\sigma}^{\dagger}d_{j\sigma}\rangle
\label{Apx9.1},
\end{equation}
\begin{equation}
\langle [A_{2\vec{k}\sigma},B_{\vec{k}j\sigma}^{(1)}]_{(+)}
\rangle=\langle n_{0-\sigma }^{d}n_{j\sigma}^{d}\rangle
-e^{i\vec{k}\cdot \vec{R}_j}\langle
d_{0\sigma}^{\dagger}n_{j-\sigma}^{d}d_{j\sigma}\rangle
\label{Apx10},
\end{equation}
\begin{equation}
\langle [A_{3\vec{k}\sigma},B_{\vec{k}j\sigma}^{(1)}]_{(+)}\rangle=0
\label{Apx11},
\end{equation}
\begin{equation}
\langle [A_{4\vec{k}\sigma},B_{\vec{k}j\sigma}^{(1)}]_{(+)}\rangle=
-\langle n_{j\sigma }^{d}d_{0\sigma}^{\dagger}
d_{0-\sigma}^{\dagger}\rangle
\label{Apx12},
\end{equation}
\begin{equation}
\langle [A_{5\vec{k}\sigma},B_{\vec{k}j\sigma}^{(1)}]_{(+)}\rangle=0
\label{Apx13}
\end{equation}
where, it has been assumed that the brackets are real and
unchanged when the indices 0 and $j$ are interchanged. Also, due
to translational invariance of the system,
$n_{0\sigma}^d=n_{j\sigma}^d$. Considering the relations given by
Eqs. (\ref{Apx1}) and (\ref{Apx4}) with the results from Eq.
(\ref{Apx9.1}) to Eq. (\ref{Apx13}), the correlation function
$\langle n_{j\sigma }^{d}n_{0\sigma}^{d}\rangle$ can be written
as:
\begin{eqnarray}
\langle n_{l\sigma }^{d}n_{0\sigma}^{d}\rangle&=&\alpha_{\sigma}n_{\sigma}^d
-\alpha_{j\sigma}n_{0j\sigma}^d + \beta_{\sigma}\langle n_{0-\sigma }^{d}n_{j\sigma}^{d}\rangle
-\beta_{j\sigma}m_{j\sigma}\nonumber \\
& &+ ~\beta_{\sigma}^{(1)}\langle n_{j\sigma }^{d}d_{0-\sigma}^{\dagger}
d_{0\sigma}^{\dagger}\rangle
\label{Apx14}
\end{eqnarray}
where $n_{0\sigma}^d=n_{\sigma}^d$.
In Eq. (\ref{Apx14}), it has been introduced the following definitions:
\begin{equation}
n_{0j\sigma}^d=\langle d_{0\sigma}^{\dagger}d_{j\sigma}\rangle=\frac{1}{L}
\sum_{\vec k}{\cal F}_{\omega}G_{\vec{k}\sigma}^{11}e^{i\vec{k}\cdot \vec{R}_j}
\label{Apx15},
\end{equation}
\begin{equation}
m_{j\sigma}=\langle d_{0\sigma}^{\dagger}n_{j-\sigma }^{d}d_{j\sigma}\rangle=\frac{1}{L}
\sum_{\vec k}{\cal F}_{\omega}G_{\vec{k}\sigma}^{12}e^{i\vec{k}\cdot \vec{R}_j}
\label{Apx16},
\end{equation}
\begin{equation}
\alpha_{j\sigma}=\frac{1}{L}
\sum_{\vec k}{\cal F}_{\omega}\widetilde{G}_{\vec{k}\sigma}^{11}e^{i\vec{k}\cdot \vec{R}_j}
\label{Apx17},
\end{equation}
\begin{equation}
\beta_{j\sigma}=\frac{1}{L}
\sum_{\vec k}{\cal F}_{\omega}\widetilde{G}_{\vec{k}\sigma}^{12}e^{i\vec{k}\cdot \vec{R}_j}
\label{Apx18}
\end{equation}
and
\begin{equation}
\beta_{j\sigma}^{(1)}=\frac{1}{L}
\sum_{\vec k}{\cal F}_{\omega}\widetilde{G}_{\vec{k}\sigma}^{14}e^{i\vec{k}\cdot \vec{R}_j}
\label{Apx19}
\end{equation}
where $G_{\vec{k}\sigma}^{11}$ is given in Eq. (\ref{eq2.22}). The remaining Green's functions
$G_{\vec{k}\sigma}^{12}$ and $G_{\vec{k}\sigma}^{14}$ are given respectively by
\begin{equation}
G_{k\sigma }^{12}(\omega)=\frac{n_{-\sigma}^d(\omega - E_{55})\left[B\left(\omega\right)-
(\omega + E_{11}){\overline{\gamma}_{k}}^2\right]}
{\overline{D}\left( \omega \right)}
\label{Apx19.1}
\end{equation}
and
\begin{eqnarray}
G_{k\sigma }^{14}(\omega)&=&(n_{-\sigma}^d)^2(1-n_{-\sigma}^d)^2U\overline{\gamma}_{k}\nonumber\\
& &\times\frac{(\omega - E_{55})(\omega + E_{11}-Un_{-\sigma}^d)}
{\overline{D}\left( \omega \right)}
\label{Apx19.2}
\end{eqnarray}
where $B\left(\omega\right)$ is defined in Eq. (\ref{eq4.5}) and $\overline{D}\left( \omega \right)$
in Eq. (\ref{eq2.33}).
It is also necessary to define
\begin{equation}
\widetilde{G}_{\vec{k}\sigma }^{11}(\omega)=\frac{G_{\vec{k}\sigma}^{11}(\omega) -
G_{\vec{k}\sigma}^{12}(\omega)}{1-n_{-\sigma}^d},
\label{Apx19.3}
\end{equation}
\begin{equation}
\widetilde{G}_{\vec{k}\sigma }^{12}(\omega)=\frac{G_{\vec{k}\sigma}^{12}(\omega) -
n_{-\sigma}^dG_{\vec{k}\sigma}^{11}(\omega)}{n_{-\sigma}^d(1-n_{-\sigma}^d)}
\label{Apx19.4}
\end{equation}
and
\begin{equation}
\widetilde{G}_{\vec{k}\sigma }^{14}(\omega)=\frac{G_{\vec{k}\sigma}^{14}(\omega) -
n_{-\sigma}^dG_{\vec{k}\sigma}^{13}(\omega)}{n_{-\sigma}^d(1-n_{-\sigma}^d)}
\label{Apx19.5}
\end{equation}
where $G_{\vec{k}\sigma}^{13}$ is given in Eq. (\ref{eq3.6}).

The correlation function $\langle n_{0-\sigma }^{d}n_{j\sigma}^{d}\rangle$ present in Eq. (\ref{Apx14}),
can be obtained by repeating the procedure above using the operator $B_{\vec{k}j\sigma}^{(2)}$
(given by Eq. (\ref{Apx3})). Thus,
\begin{eqnarray}
\langle n_{j-\sigma }^{d}n_{0\sigma}^{d}\rangle&=&\alpha_{\sigma}n_{-\sigma}^d
+\beta_{\sigma}\langle n_{0-\sigma }^{d}n_{j-\sigma}^{d}\rangle +
\alpha_{j\sigma}^{(1)}n_{0j\sigma}^{(1)}\nonumber\\
& &  + \beta_{j\sigma}^{(1)}m_{j\sigma}^{(1)}
-\beta_{\sigma}^{(1)}\langle n_{j-\sigma }^{d}d_{0\sigma}^{\dagger}
d_{0-\sigma}^{\dagger}\rangle
\label{Apx19.6}
\end{eqnarray}
where
\begin{equation}
n_{0j\sigma}^{(1)}=\langle d_{j-\sigma}d_{0\sigma}\rangle=\frac{1}{L}
\sum_{\vec k}{\cal F}_{\omega}G_{\vec{k}\sigma}^{13}e^{i\vec{k}\cdot \vec{R}_j}
\label{Apx20},
\end{equation}
\begin{equation}
m_{j\sigma}^{(1)}=\langle d_{j-\sigma}n_{j\sigma }^{d}d_{0\sigma}\rangle=\frac{1}{L}
\sum_{\vec k}{\cal F}_{\omega}G_{\vec{k}\sigma}^{14}e^{i\vec{k}\cdot \vec{R}_j}
\label{Apx21},
\end{equation}
and
\begin{equation}
\alpha_{j\sigma}^{(1)}=\frac{1}{L}
\sum_{\vec k}{\cal F}_{\omega}\widetilde{G}_{\vec{k}\sigma}^{13}e^{i\vec{k}\cdot \vec{R}_j}
\label{Apx22}
\end{equation}
with
\begin{equation}
\widetilde{G}_{\vec{k}\sigma }^{13}(\omega)=\frac{G_{\vec{k}\sigma}^{13}(\omega) -
G_{\vec{k}\sigma}^{14}(\omega)}{1-n_{-\sigma}^d}.
\label{Apx21.1}
\end{equation}

Reversing the spin labels i.e., $\sigma \rightarrow -\sigma$ in
Eq. (\ref{Apx19.6}) and substituting the result into Eq.
(\ref{Apx14}), then
\begin{eqnarray}
\langle n_{j\sigma }^{d}n_{0\sigma}^{d}\rangle&=&\frac{\alpha_{\sigma}n_{\sigma}^d
-\alpha_{j\sigma}n_{0j\sigma}^d + \beta_{\sigma}\alpha_{-\sigma}n_{\sigma}^d
-\beta_{j\sigma}m_{j\sigma}}{1-\beta_{\sigma}\beta_{-\sigma}}\nonumber \\
& &+ \frac{1}{1-\beta_{\sigma}\beta_{-\sigma}}
\left [\beta_{\sigma}(\alpha_{j-\sigma}^{(1)}n_{0j-\sigma}^{(1)}
+\beta_{j-\sigma}^{(1)}m_{j-\sigma}^{(1)})\right.\nonumber \\
& &\left.+(\beta_{\sigma}\beta_{-\sigma}^{(1)}
+\beta_{\sigma}^{(1)})\langle n_{j\sigma }^{d}d_{0-\sigma}^{\dagger}
d_{0\sigma}^{\dagger}\rangle \right ]
\label{Apx23}.
\end{eqnarray}

For evaluate the two last correlation functions present in Eq. (\ref{eq4.8}),
the following operators have been introduced
\begin{equation}
B_{\vec{k}j\sigma}^{(3)}=\frac{1}{\sqrt{L}}\sum_ie^{-i\vec{k}\cdot\vec{R}_i}
d_{i+j\sigma}^{\dagger}d_{i+j-\sigma }d_{i-\sigma }^{\dagger}
\label{Apx24}
\end{equation}
and
\begin{equation}
B_{\vec{k}j\sigma}^{(4)}=\frac{1}{\sqrt{L}}\sum_ie^{-i\vec{k}\cdot\vec{R}_i}
d_{i+j\sigma}^{\dagger}d_{i+j-\sigma }^{\dagger}d_{i-\sigma }
\label{Apx25}.
\end{equation}
Using $B_3$, and following the procedure outlined above, the correlation function
$\langle d_{j\sigma }^{\dagger}d_{j-\sigma }d_{0-\sigma }^{\dagger}d_{0\sigma }\rangle$
is given by
\begin{eqnarray}
\langle S_jS_0\rangle&=&\langle d_{j\sigma }^{\dagger}d_{j-\sigma }d_{0-\sigma }^{\dagger}d_{0\sigma }\rangle=
-\frac{1}{1+\beta_{\sigma}}\left[
\alpha_{j\sigma}n_{0j-\sigma}^d\right. \nonumber \\ 
& & \left.+ \beta_{j\sigma}m_{j-\sigma}
-\alpha_{j\sigma}^{(1)}n_{0j-\sigma}^{(1)} - \beta_{j\sigma}^{(1)}m_{j-\sigma}^{(1)}\right].
\nonumber \\
\label{Apx26}
\end{eqnarray}
Similarly, using $B_4$
\begin{eqnarray}
\langle d_{j\sigma }^{\dagger}d_{j-\sigma }^{\dagger}d_{0-\sigma }d_{0\sigma }\rangle&=&
\frac{\alpha_{j\sigma}n_{0j-\sigma}^d +\beta_{j\sigma}(n_{0j-\sigma}^d -m_{j-\sigma} ) }
{1-\beta_{\sigma}}\nonumber \\
& &+\frac{\beta_{\sigma}^{(1)}\langle n_{0\sigma }^dd_{j\sigma }^{\dagger}d_{j-\sigma }^{\dagger}
\rangle}{1-\beta_{\sigma}}
\label{Apx27}
\end{eqnarray}
where the $d$-wave symmetry has been considered, therefore, $\langle
d_{j\sigma }^{\dagger}d_{j-\sigma }^{\dagger}\rangle=0$.

The four $B^{(p)}$ operators introduced up to now are exactly the
same operators used by Roth in Ref. \cite{ref8} to obtain the band
shift $W_{k\sigma}$ in the normal state and without hybridization.
However, in the present work, due to the presence of the hole
operators (see Eq. (\ref{eq2.9})), a new $B$ operator, which is
given by
\begin{equation}
B_{\vec{k}j\sigma}^{(5)}=\frac{1}{\sqrt{L}}\sum_ie^{-i\vec{k}\cdot\vec{R}_i}
d_{i\sigma}^{\dagger}d_{i+j\sigma }d_{i+j-\sigma }
\label{Apx28},
\end{equation}
has been introduced. With this operator, the correlation function
$\langle n_{0\sigma }^dd_{j\sigma }^{\dagger}d_{j-\sigma }^{\dagger}
\rangle$ present in Eq. (\ref{Apx27}) may be evaluated. Thus,

\begin{eqnarray}
\langle d_{0\sigma}^{\dagger}d_{j-\sigma }d_{j\sigma }d_{0\sigma}\rangle&=&
\frac{\alpha_{j\sigma}^{(1)}n_{0j\sigma}^d +\beta_{j\sigma}^{(1)}(n_{0j\sigma}^d -m_{j\sigma} ) }
{1-\beta_{\sigma}}.\nonumber \\
\label{Apx29}
\end{eqnarray}
Substituting the result (\ref{Apx29}) into Eq. (\ref{Apx27}), the
correlation function $\langle d_{j\sigma }^{\dagger}d_{j-\sigma
}^{\dagger}d_{0-\sigma }d_{0\sigma }\rangle$ can be rewritten as:
\begin{eqnarray}
\langle d_{j\sigma }^{\dagger}d_{j-\sigma }^{\dagger}d_{0-\sigma }d_{0\sigma }\rangle&=&
\frac{\alpha_{j\sigma}n_{0j-\sigma}^d +\beta_{j\sigma}(n_{0j-\sigma}^d -m_{j-\sigma} ) }
{1-\beta_{\sigma}}\nonumber \\
& &+\frac{\beta_{\sigma}^{(1)}[\alpha_{j\sigma}^{(1)}n_{0j\sigma}^d
+\beta_{j\sigma}^{(1)}(n_{0j\sigma}^d -m_{j\sigma} )]}{(1-\beta_{\sigma})^2}.\nonumber \\
\label{Apx30}
\end{eqnarray}

The result given in Eq. (\ref{Apx29}) can be used in Eq. (\ref{Apx23}) to obtain
\begin{eqnarray}
\langle n_{j\sigma }^{d}n_{0\sigma}^{d}\rangle&=& (n_{0j\sigma}^d)^2 -
\frac{\alpha_{j\sigma}n_{0j\sigma}^d + \beta_{j\sigma}m_{j\sigma}}
{1-\beta_{\sigma}\beta_{-\sigma}}\nonumber \\
& &+ \frac{1}{1-\beta_{\sigma}\beta_{-\sigma}}
\left [\beta_{\sigma}(\alpha_{-\sigma}^{(1)}n_{0j-\sigma}^{(1)}
+\beta_{j-\sigma}^{(1)}m_{j-\sigma}^{(1)})\right.\nonumber \\
& &\left.-\frac{\alpha_{j\sigma}^{(1)}n_{0j\sigma}^d
(\beta_{\sigma}\beta_{-\sigma}^{(1)}+\beta_{\sigma}^{(1)})}{1-\beta_{\sigma}}\right.\nonumber \\
& &-\left.\frac{\beta_{j\sigma}^{(1)}(n_{0j\sigma}^d-m_{j\sigma})
(\beta_{\sigma}\beta_{-\sigma}^{(1)}+\beta_{\sigma}^{(1)})}
{1-\beta_{\sigma}}\right ]
\label{Apx31}.
\end{eqnarray}

Finally, with the results (\ref{Apx29}), (\ref{Apx30}) and (\ref{Apx31}) into Eq. (\ref{eq4.9}),
the following result has been obtained
\begin{equation}
n_{-\sigma}^d(1-n_{-\sigma}^d)W_{\vec{k}\sigma }^d=h_{1\sigma}
+\sum_{j\neq 0}t_{0j}^{d}e^{i\vec{k}\cdot\vec{R}_j}(h_{2j\sigma}
+h_{3j\sigma})
\label{Apx32}
\end{equation}
where
\begin{equation}
h_{1\sigma}= -\sum_{j\neq 0}t_{0j}^{d}(n_{j0\sigma }^{d} - 2m_{j\sigma })
\label{Apx33},
\end{equation}
\begin{eqnarray}
h_{2j\sigma}&=&-\left\{\frac{\alpha_{j\sigma}n_{0j\sigma}^d + \beta_{j\sigma}m_{j\sigma}}
{1-\beta_{\sigma}\beta_{-\sigma}} + \frac{\alpha_{j\sigma}n_{0j-\sigma}^d +
\beta_{j\sigma}m_{j-\sigma}}{1+\beta_{\sigma}}\right.\nonumber\\
& &+\left.\frac{\alpha_{j\sigma}n_{0j-\sigma}^d +\beta_{j\sigma}(n_{0j-\sigma}^d
-m_{j-\sigma} ) }{1-\beta_{\sigma}}\right\},
\label{Apx34}
\end{eqnarray}
\begin{eqnarray}
 h_{3j\sigma} &=&\phi_{j\sigma}\left\{\frac{\alpha_{j\sigma}^{(1)}n_{0j\sigma}^d
+\beta_{j\sigma}^{(1)}(n_{0j\sigma}^d -m_{j\sigma} )}{1-\beta_{\sigma}}\right\}\nonumber\\
& &- \frac{\beta_{\sigma}[\alpha_{j-\sigma}^{(1)}n_{0j-\sigma}^{(1)}
+\beta_{j-\sigma}^{(1)}m_{j-\sigma}]}{1-\beta_{\sigma}\beta_{-\sigma}}\nonumber\\
& & -\frac{\alpha_{j\sigma}^{(1)}n_{0j-\sigma}^{(1)} + \beta_{j\sigma}^{(1)}
 m_{j-\sigma}^{(1)}}{1+\beta_{\sigma}}
\label{Apx35}
\end{eqnarray}
with
\begin{equation}
\phi_{j\sigma} =\frac{(\beta_{\sigma}^{(1)}}{1-\beta_{\sigma}}
+\frac{\beta_{\sigma}\beta_{-\sigma}^{(1)}+\beta_{\sigma}^{(1)}}{1-\beta_{\sigma}\beta_{-\sigma}}
\label{Apx36}.
\end{equation}



\end{document}